\documentstyle[11pt,newpasp,twoside]{article}
\reversemarginpar

\begin{document}
\title{LiBeB Production  and Associated Astrophysical  Sites}
\author{Elisabeth Vangioni-Flam}

\affil{Institut d'Astrophysique de Paris, 98 bis Bd Arago, 75014 Paris, France}
\author{Michel Cass\'e}
\affil{ Service d'Astrophysique, Orme des Merisiers, CEA, 91191 Gif sur Yvette, France}

\begin{abstract}
The various modes of spallative LiBeB production are summarized, and classified
 according to their dependence or independence on the abundance of medium
 heavy elements (CNO) illustrated by that of oxygen in the interstellar medium.
 The predictions of the models are confronted to the available observational correlations
 (Be, B vs O). Clearly, a primary mechanism should lead to a slope one in the lg(Be/H) vs [O/H]
 plot and a secondary mechanism to a slope two. Due to the 
 ambiguity of the O data, another criterion, based on energetics, can help us to select an adequate
 model. A purely secondary origin in the very early Galaxy is much more 
 energy demanding than a primary one. Indeed, 
 magnesium seems to be a possible surrogate of oxygen and iron
 since i) it is 
spectroscopically  more easy to cope with and ii)  its nucleosynthetic
  yield is independent of the mass cut and does not depend on metallicity.
\end{abstract}

\section{Introduction}
A turning point in the theory of the origin and evolution of light elements 
has been the observation of a linear relationship between both Be and B and 
Fe in metal poor halo stars (Rebolo, Molaro \& Beckman 1988; Duncan et al 1992, 1997;
 Gilmore et al 1992, Boesgaard \& King 1993; Ryan, Norris, Bessel \& Deliyannis 1994, Primas et al 1999).
 But recently, the debate has taken a complex turn, due to a modification
 of the O-Fe correlation indicated by the data of Israelian et 
al (1998) and Boesgaard et al (1999). This revision is however not universally 
admitted, and the debate is still open (e.g. Fulbright \& Kraft 1999; 
Gustafsson 1999; Reetz 2000). Since the last International Cosmic Ray 
Conference (Barring 2000) and the most recent LiBeB meeting
 (Ramaty et al 1999), the situation has not ceased to
 evolve.

 Prior to the proposal of change of the O-Fe relation,
 the situation was the following: in order to explain the observed proportionality
 of  the Be/H ratio to Fe/H (itself taken proportional to O/H in halo stars as 
implied at that time by the data and by the SNII calculated yields), 
it was necessary to invoke a primary production
 mechanism (i.e. a production rate of LiBeB independent of the interstellar metallicity),
 driven by the break up in flight of C and O colliding with H and He
 in the ISM (for reviews see Vangioni-Flam et al 1999b, 2000a,b).

 Now 
the reality of this primary component is questioned on the basis of the "new" O-Fe
 correlation. Concerning oxygen, however, the situation is not settled. The abundances 
derived from the forbidden (OI) line, which is certainly the most accurate 
source when it is not too weak, suggest a plateau but measurements of near 
IR OH band in dwarfs and subgiants lead to a rising trend with decreasing 
Fe/H. In contrast, Fulbright \& Kraft (1999) have analysed in great details 
the (OI) spectral region in the two metal poorest stars of the Israelian et al 
sample and have found a lower O/Fe ratio. As stressed for instance by Pagel 
(1999), both methods, indeed, have their drawbacks and technical difficulties: 
the OH bands are subject to uncertainties in UV continuum absorption 
(Balachandran \& Bell 1998) and effective temperature, while the forbidden lines
 are so weak at low metallicity that the determination of the continuum becomes problematic.

 So the situation is wide open and one is inclined to propose a different metallicity
 index,  less ambiguous than O. The closer element to O whose abundance
 is widely measured is Mg. It has been chosen, due to its various advantages,
 both observational  and theoretical, as the reference element to follow
 galactic evolution by Thomas et al (1998),  Fuhrmann (1998) and  Shigeyama 
 \& Tsujimoto (1998). Indeed i) Mg seems easier to measure than O and ii
) Mg and O are coproduced in SNII explosions, which are the main sources of these
 elements (Woosley \& Weaver 1995, Thielemann, Nomoto \& Hashimoto, 1996).  
Moreover the Mg yield is independent of the mass cut, and does not significantly
 depend on metallicity (Umeda et al 2000). In the light of existing data,
 the Mg/Fe vs Fe/H correlation is rather flat up to Fe/H = -1, like that of
 other alpha elements (Mc Williams 1997, Pagel \& Tautvaisiene, 1995).
 Taken at face value, (assuming Mg proportional to O, on
 nucleosynthetic grounds, neglecting the peculiar behaviour of the most massive stars which are 
 marginal in the chemical evolution budget), 
these data indicate the need of a primary component, then we are back to
 the previous situation (Cass\'e et al 1995; Vangioni-Flam et al 1996,
  1998; Ramaty et al 1996).
 Thus, on this sole basis,
 a purely secondary origin of Be in the halo (Fields \& Olive 1999a) 
driven by the standard GCR seems inadequate. However, the situation is not completely settled.

  The outline of this paper is the following: in section 2
 we recall the basic production mechanisms of LiBeB,
 in section 3 we decline the astrophysical agents and sites, in section 4,
  we compare the various models designed,
 in section 5 we propose key observations to remove the present ambiguities.

\section{Nucleosynthesis of LiBeB}
 
\subsection{Thermal production and destruction}

 Thermal nucleosynthesis in the Big Bang produces negligible amounts of Be and B.
 Only $^7$Li is synthesized in significant amounts.
  Moreover, LiBeB do not survive stellar temperatures except in a thin surface layer where 
they are observed, reflecting the interstellar composition inherited by the star at birth.
 $^7$Li is however thought to be produced by AGB stars (Abia et al 1993) and novae 
 (Hernanz et al 1996) and also by SNII through neutrino spallation of carbon.
  This neutrino spallation in carbon shells of type II supernovae is expected to 
 produce also $^{11}$B, but the yields are sensitive to the assumed 
 temperature (energy) of the neutrinos, which is uncertain (see Hartmann et 
 al 1999 for a review). 

\subsection{Non Thermal production}
 Nuclear spallation, i.e. the break up of medium
 heavy elements by collisions with protons and alphas remains the leading 
production process of light elements in the cosmos (Meneguzzi, Audouze \& Reeves 1971). In principle, 
 all isotopes of interest are generated either by the interaction of fast p 
and alphas on CNO at rest in the ISM, or conversely  by the interaction of 
 fast C and O (principally) with ambient H and He,
 supplemented by the alpha +alpha reaction giving rise exclusively to Li isotopes.
 (e.g. Reeves 1994). 

The cross sections are well measured (Read \& 
Viola 1985; Webber et al 1990 a, b), and have been updated recently by Ramaty 
et al (1997). The hierarchy of the cross sections reflects that of the abundances 
of  the light nuclei in nature. In a collision between a proton and an oxygen nucleus, the probability 
of production of $^{11}$B is higher than that of  $^{10}$B which is itself 
higher than that of $^9$Be. Thus, it is not surprising that the abundances of these 
three isotopes go in declining order. This is a stricking example of a
 direct application of nuclear physics in the understanting of natural abundances. 
Note that the peaks of the cross sections lie at low energy especially that
  of alpha + alpha (Read \& Viola 1985). Thus low energy particles (about 10 MeV/n) have to be inserted 
carefully in the treatment of the problem.  Note also that at low energy, 
where the alpha + alpha reaction is operating at full strength, $^7$Li and $^6$Li are produced
 in comparable amounts (1.5), which is at variance with the $^7$Li/$^6$Li ratio observed in
 meteorites (12.5). A stellar source of pure $^7$Li has to be invoked  to explain
 this high value (see above). 

The production rate, in the most general case, is function i) of the number 
density of the target nuclei ($N_T$)  and their composition, ii) of the flux of the projectiles ($\Phi$)
and iii) of the cross section averaged over the energy spectrum in the interaction region :

           $dN(L)/dt = N_T<\sigma>\Phi$. 

If p and alpha 
are the projectiles and CNO the targets, one deals with a "secondary" production,
 thus:

    $dN(L)/dt = N_{CNO}<\sigma>\Phi_{p,\alpha}$.
 
It is assumed (quite reasonably) that the flux of energetic particles is 
proportional to the supernova rate since they are thought to be
 the main agents of acceleration  through the shock 
waves they produce. Due to the fact that SNII are also the main O producers, one expects that:

  $\Phi(t) prop. d(SN(t))/dt prop. d(O/H)/dt$.

  On the other hand, the CNO abundance cumulated in the ISM up to time t, is 
proportional to the total number of SN having exploded from 0 to t. 
Summarizing:

 $d(Be/H)/dt prop. (O/H)d(O/H)/dt$ or after integration Be/H prop. $(O/H)^2$.

In contrast, if C, O are the projectiles and H, He the targets,
a primary production arises governed by the equation:

   $dN(L)/dt = N_{HHe}<\sigma>\Phi_{CO}$

 Now with the same hypothesis than above, and considering that the target
 abundances (H, He) do not evolve significantly:

  $d(Be/H)/dt prop. d(O/H)/dt and Be/H prop. O/H$

\section{Astrophysical agents and sites}
\subsection{Galactic Cosmic Rays (GCR)}
The energy spectrum of GCR is directly observed above, say, 1GeV/n. Below, it is deduced from various indirect
 observations (Strong \& Moskalenko 1999).
 It is reasonably well explained by the diffuse
 shock wave acceleration mechanism (Blandford \& Ostriker 1978; Jones 
\& Ellison 1991; Ellison et al 1997). The observed composition is 
 extrapolated back 
to the sources thanks to a classical propagation model. 

It is edifying to compare
 the elemental and isotopic source composition to that of other 
materials of known abundances (the solar system for instance, or that of the 
supernova ejecta computed with stellar models). Indeed SNIa alone do not
 fit O/Fe, Ne/Fe, Mg/Fe neither the s process elements, whereas,   SNII alone do 
not fit Fe/Co/Ni, neither the s process elements. (see e.g.  Meyer 1996, 2000). 
 Indeed, the
 nucleosynthetic origin of  the groups (Mg, Si, Ca, Fe, Ni), (Sr, Zr, Ba, Ce), (Pt 
peak, actinides) are all different (explosive burning,
 s-process and r process, respectively) and their production sites are also different,
 thus the solar mix is a complex 
mixture of all that, historitically built up,  and individual sources
 are unlikely to lead to the GCR source abundances
 (since they are similar to that of the solar system). Note that in this example we
 have chosen only refractory elements that are not supposed to be affected
 by selective effects (see below).

A stricking fact, of great importance for our purpose, is the similarity 
between the isotopic composition of the GCRs and of the solar system (Connell \& Simpson 1997;
 Stone et al 1998; Wiedenbeck et al 1999),
 which is a mixture of the products of generations and generations of stars
 of different masses indicating  at face value that GCR are nuclei accelerated
 out of a normal reservoir and not strange, exotic, stars or objects.
 
In recent years, grains have been
 considered to play a central role in explaining the pecularities of the
 GCR source composition (Meyer et al 1997; Ramaty et al 1997).
 It is assumed that grain debris are more efficiently accelerated than
 elements in the gas phase and  grain models have 
progressively replaced the traditional two step acceleration mechanism 
(injection by flare stars and acceleration by supernova shocks) in which 
grains were undesirable (Cass\'e \& Goret 1978, Meyer 1985; Silberberg \& Tsao 1990). 

However the situation is not absolutely settled (Shapiro 1999, 
Silberberg et al 2000; Cass\'e \& Vangioni-FlamÚ 2000, in 
 preparation). Anyway, among  the  grain
 supporters themselves, there are divergent views concerning the origin
 and nature of the grains of interest. In one camp, they are supposed 
interstellar (Meyer and coworkers) and on the other they are supposed
 to contain fresh products of nucleosynthesis (Ramaty and colleagues). 
This has a strong bearing on the primary or secondary character 
of the LiBeB production mechanism. If the grains that are impacted by
 the accelerating shock waves have the ISM composition, then, the resulting CR composition
 should reflect that of the ISM which is the true reservoir of CR particles and, as it is 
 H, He dominated, the process is secondary.
  
 Ramaty and coworkers (see Ramaty et al 2000
 and references therein) assume that GCR originate from grains loaded
 with freshly synthesized nuclei, C and O that are released by SNII and accelerated by
 shock waves in galactic superbubbles (SB). Subsequently, they interact
 with the surrounding interstellar medium to give LiBeB in a primary way.  
However, this proposal has to face the following objections: in the 
superbubble (SB), high temperature context,  according to the observed trend 
(Cardelli 1996; Savage \& Sembach, 1996; Jenkins et al 1998; Howk, Savage \& Fabian 
1999), grains have little chances to remain intact. Only their refractory
 cores would survive in the hot phase. Silicon, for instance would be significantly  
evaporated (see fig 5 in Savage \& Sambach 1996) with respect to Fe, and 
the Si/Fe ratio in the accelerated particles should be different than solar 
contrary to what is deduced from the GCR observations. Thus grains in SB 
are unlikely to account, in a detailed manner, for the present GCRS composition.
 Moreover, the $^{34}$S/$^{32}$S ratio, derived at the cosmic ray sources poses also a problem
 (Cass\'e \& Vangioni-Flam 2000 in preparation). Finally, the grain mechanism does
 not go without gas (H, He) acceleration and thus particle accelerated in superbubbles
 cannot induce a purely primary process all time long and as the Galaxy evolves, the secondary
  process becomes more and more important. In this case, the transition primary-secondary
 will take place at about [Fe/H] about -1 (taking an average superbubble metallicity of about
 5 times solar, Higdon et al 1998). Thus a purely primary component seems unlikely.

On this basis we are tempted to conclude that standard GCR, as traditionally thought, 
act as a secondary source of LiBeB, and if a primary component is made necessary
 by the data, it should be 
different from the standard GCR one. It could come from SB's under the condition that
 this SB component is confined to low energy not to spoil the observed CGR composition.

\subsection{Superbubble Accelerated Particles (SAP)}
 Superbubbles gather a great number of massive stars which explode 
as type II supernovae, enriching the surrounding medium in fresh products of 
nucleosynthesis, and among them, oxygen. The shock wave and turbulence 
sutained by a given supernova, accelerate the material enriched by the previous
 ones (Bykov \& Fleisshman 1992; Bykov 1995; Bykov et al 2000). 
The energy spectrum obtained could depart significantly from the GCR one, 
depending on the detail of the mechamism and more precisely on the escape 
time of the fast particles, which is poorly known (Klepach et al 1999).
As said previously, an energy cut off should be imposed to avoid 
contamination of the (observed) GCR abundances. This energy cut off
 is still a free parameter and  we can  choose it in order to avoid 
energetic problems (Ramaty et al 1996), i.e. in the range 30-300 MeV/n. 
Then, all in all, superbubbles appear to be the best agents of a primary 
production process of LiBeB (Parizot \& Drury 1999; Bykov et al 2000). Admittedly, this 
theoretical proposal remains to be firmly (experimentally) substantiated by X 
ray and gamma-ray line observations (see the proceedings edited by Ramaty et al 
1999 for a general discussion). The search for a non thermal low energy 
component as the one once claimed to be discovered by the COMPTEL 
experiment  but discarded afterwards (Bloemen et al 1999) is, in our minds, one 
of the major objectives of the European INTEGRAL satellite to be lauched in 2001.
 For the time being, we can only speculate on the composition and the 
spectrum of this hypothetical (but somewhat necessary) component. In the halo phase, the 
composition is taken as representative of the ejecta of massive low 
metallicity stars. It is highly enriched in O w.r.t present CRs.
(Woosley \& Weaver 1995). This composition however is expected to
 vary in time due to metallicity dependent mass loss rate (Vangioni-Flam et al 1997).
 
\section{Production and evolution of LiBeB}
\subsection{Observed correlations}
Turning to observations, we are confronted to a certain ambiguity (as 
mentionned in the introduction). According to the choice of the conversion 
between Fe and O abundances we get different conclusions. Relying on the 
Israelian et al (1998) and Boesgaard et al (1999) relation  we get 
Be proportional to  $O^{1.7}$, i.e. nearly secondary, but taking the conventional 
O-Fe relation, strengthened by the Mg data, we get Be proportional to O, i.e. 
purely primary, at least in the halo phase. Thus the O/Fe behaviour at low 
metallicity is a central issue, once again. Furthermore non LTE effects on B 
(Kiselman 1999)  make the situation even more
 complex (see also Primas 2000).
 Thus we must rely on a independent argument to answer the question: is a 
primary component really necessary? This argument of energetic nature has 
been essentially developed by Ramaty et al (1996, 1997, 2000). 
To briefly summarize: it seems that a primary component is 
required by both the Mg-Fe relation (more secure than O-Fe) and by energetic arguments.

\subsection{The three evolutionary models}
Concerning the galactic evolutionary 
models, there exists three types of them
i) a pure secondary 
standard GCR based on the variable  O/Fe ratio (Fields \& Olive 1999a)  
ii) a pure primary GCR from SB, which, suprisingly is still valid with flat
 or variable O/Fe (Ramaty et al 2000) iii) an hybrid model 
(Standard GCR + SAP based on a flat O/Fe ratio (Vangioni-Flam et al 1998)
 or based on a variable O/Fe ratio (Fields et al 2000).

\subsection{Energetic requirements}
The number of atoms of Be produced per erg injected per supernova is 
promoted to the role of a selection criterion for the theoretical models (both 
primary and secondary). Stated differently, are there enough SNII  and  are 
they sufficiently efficient to produce all the Be observed in halo stars in the 
primary and/or secondary cases? According to Ramaty et al (2000) the 
pure secondary standard GCR produce BeB at high energy cost in the early 
Galaxy. SAP or primary GCR are much more economical. 
 Thus a very plausible solution is that a primary component appears first, and 
a secondary component takes over afterwards, at a metallicity (O/H ratio) 
which remains to be determined precisely. The position of the break depends
 on the Fe-O correlation used. Relying on the new one, and using the 
analysis based on the IRFM data, (Fields et al 2000) one finds a break point at about
  [O/H] = -1.6. On 
the other hand, if the ancient correlation is chosen, motivated by the 
magnesium data, the break point (if any) is at higher [O/H] (about -1). Anyway, the existence
 of a primary component in the early stages of the 
evolution of the Galaxy seems mandatory, irrespective of the abundance data 
used, and this is a strong conclusion. Definitively, a primary component is 
required to fulfill the energetic constraint. What is the nature of this 
primary component? In our opinion, once again, it is distinct from GCR
 (assumed by Ramaty et al 2000 to originate from SBs), 
since, as said previously, we do not think that it is possible to  identify SAP 
and GCR, as Higdon et al (1998) did, due to their different inferred 
composition. 

Thus, we support the view that a low energy component
 is at work, complementing GCR to produce its lot of LiBeB, specifically in the halo phase.
  Is this component still active?
 Nuclear gamma ray line astronomy will say. Thus our hopes are 
related to the INTEGRAL satellite. Our conclusion is that an hybrid model 
combining both a primary and a secondary components reconcile the 
abundance observations and energy requirement. The two components  
 dominate sequentially. Of course the primary component, related to SAP 
 would be overwhelming in the halo phase and is afterwards dominated by the secondary one, which 
constitutes the standard GCR. The SAP component plays a major role in the LiBeB  production 
when the galactic gas is almost devoid of medium elements i.e. in the early Galaxy. 
Now, what is the ultimate reservoir of GCR? Grains in the ISM or stellar surfaces?
 This point is left to a future discusssion.

\section{Conclusion}
 Ambiguities on abundance data preclude definitive conclusion: is O/Fe flat 
or not at low Z? For the time being, the answer to this question depends 
on the observer to whom it is posed. Hopefully the debate will 
clarify in the next years. In the mean time we propose to rely  on magnesium 
which is used as a secure metallicity index by a growing number of people. 
Special care should be taken to make NLTE corrections on Fe and B at low Z.
 A primary process probably is made necessary by energetic requirements. We
 endorse the view that two different components are responsible for 
the synthesis and evolution of LiBeB in the Galaxy. This hybrid model 
invoking the operation of both GCR (extracted by flares from stellar surfaces 
and/or grain debris in the ISM) subsequently accelerated by shock waves and 
a primary component of lower energy coming probably from superbubbles  
fulfil  (or at least do not violate) all the composition constraints on i) the 
present cosmic radiation ii) the light element abundances in stars of all 
metallicities, including $^6$Li (Vangioni-Flam et al 1999a; Fields \& Olive 1999b).

 What to do next? Obviously 
measure. One would like idealy to get simultaneously the abundances of
 $^{6,7}$Li, Be, B, O, Mg and Fe in the same star, this for many halo members
 of various metallicities,  which does not seems out of reach of a
 dedicated VLT program (Cayrel, private communication). The solution of 
the LiBeB riddle is definitly in the hands of observers. The best way 
to reveal the presence of a low energy flux of C and O in the Galaxy 
related supperbubbles is the observation of broad gamma ray lines arising
 from their excitation and deexcitation in flight. The best hope to 
detect them in Vela, Orion and other star forming regions is offered by the 
european INTEGRAL satellite to be launched around 2001.

Aknowledgements

We thank R. Cayrel and C. Furhmann for having drawn pure attention on the 
reliability of using magnesium as a metallicity index. We thank also Andrei Bykov and
 Vladimir Ptuskin for illuminating discussions on superbubbles.

\end{document}